\documentclass[12pt,twoside]{article}
\usepackage{fleqn,espcrc1}
\input psfig

\title{Gamma-Ray Bursts\thanks{Invited talk presented at the
7th International Symposium on Particles, Strings and Cosmology
(Dec. 1999, Lake Tahoe, California).}}

\author{Eli Waxman\address{Dept. of Condensed Matter Physics, 
Weizmann Institute of Science,
Rehovot 76100, Israel;
E-mail: waxman@wicc.weizmann.ac.il}\thanks{Alon Fellow}\thanks{Incumbent of 
the Beracha foundation career development chair}
\thanks{Work supported in part 
by BSF Grant 9800343, AEC Grant 38/99 and MINERVA Grant}
}

\begin{document}

\maketitle

\begin{abstract}

Ultra-high-energy, $>10^{19}$~eV, cosmic-ray and 
high energy, $\sim10^{14}$~eV, neutrino production in GRBs is discussed in the 
light of recent GRB and cosmic-ray observations.
Emphasis is put on model
predictions that can be tested with operating and planned 
cosmic-ray and neutrino detectors, and on the prospects of testing for 
neutrino properties.

\end{abstract}

\section{Introduction}

The origin of GRBs,
bursts of 0.1 MeV---1 MeV photons lasting for a few seconds, 
remained unknown for over 20 years, primarily because GRBs 
were not detected prior to 1997 at wave-bands other than 
$\gamma$-rays \cite{Fishman}.
The isotropic distribution of bursts over the sky
suggested that GRB sources lie at cosmological distances, and
general phenomenological considerations were used to argue that the
bursts are produced by the dissipation of the kinetic
energy of a relativistic expanding fireball (see [2] for review).

Adopting the cosmological fireball hypothesis, it
was shown that the physical conditions in the fireball dissipation region 
allow Fermi acceleration of protons
to energy $>10^{20}{\rm eV}$ \cite{W95a,VMU}, and that
the average rate at which energy is emitted as $\gamma$-rays
by GRBs is  
comparable to the energy generation rate of UHECRs in a model where
UHECRs are produced by a cosmological distribution of sources \cite{W95a}.
Based on these two facts, it was suggested that GRBs and UHECRs have
a common origin (see [5] for a recent review). 

In the last two years, afterglows of GRBs have been discovered in X-ray, 
optical, and radio wave bands \cite{AG_obs_review}.
Afterglow observations confirmed the cosmological origin of the bursts,
through the redshift determination of several GRB host-galaxies, and
confirmed standard model predictions of afterglows
that result from the collision of an expanding fireball with
its surrounding medium \cite{AG_th_review}. 
These observations therefore provide strong
support for the GRB model of UHECR production.

In this review, UHECR and neutrino production in GRBs is discussed in the 
light of recent GRB and UHECR observations. 
The fireball model is briefly described in \S2.1, and proton acceleration
in GRB fireballs is discussed in \S2.2. 
Implications of recent afterglow observations to high energy particle 
production are discussed in \S3. 
Model predictions are shown to be consistent with the observed
UHECR spectrum in \S4. Predictions
of the GRB model for UHECR production, that can be tested with future 
UHECR experiments, are discussed in \S5. 
High energy neutrino production in fireballs and its implications for future
high energy neutrino detectors are discussed in \S6.

\section{UHECR from GRB fireballs}

\subsection{The fireball model}

In the fireball model of GRBs, a compact source, of linear scale
$r_0\sim10^7$~cm, produces a wind characterized by an average luminosity 
$L\sim10^{52}{\rm erg\,s}^{-1}$ and mass loss rate $\dot M=L/\eta c^2$.
At small radius, 
the wind bulk Lorentz factor, $\Gamma$, 
grows linearly with radius, until most of the wind energy is converted
to kinetic energy and $\Gamma$ saturates at $\Gamma\sim\eta\sim300$.
Variability of the source on time scale $\Delta t\approx r_0/c\sim1$~ms, 
resulting
in fluctuations in the wind bulk Lorentz factor $\Gamma$ on similar
time scale, lead to internal shocks
in the expanding fireball at a radius 
$r_i\approx\Gamma^2c\Delta t$. These shocks reconvert 
part of the kinetic energy to internal energy, which is then 
radiated as 
$\gamma$-rays by synchrotron emission of
shock-accelerated electrons. The $\gamma$-ray flux is observed 
to vary on time
scale $t_{\rm var}\sim r_i/\Gamma^2 c\sim\Delta t$. 

As the fireball expands, it drives a relativistic shock (blast wave)
into the surrounding gas. At a radius $r\sim\Gamma^2cT$, where 
$T\sim10$~s is the wind duration, most of the fireball energy 
is transferred to the surrounding gas, and
the flow approaches self-similar expansion. 
The shock driven into the ambient medium at this stage
continuously heats new gas, and accelerates relativistic 
electrons that produce by synchrotron emission 
the delayed radiation, ``afterglow,''
observed on time scales of
days to months.
As the shock-wave decelerates, the emission shifts with time to
lower frequency.

\subsection{Fermi acceleration in GRBs}

The observed GRB and afterglow radiation is produced
by synchrotron emission of shock accelerated
electrons. In the region where electrons are accelerated, 
protons are also expected to be
shock accelerated. This is similar to what is thought to occur in supernovae 
remnant shocks \cite{Bland87}. Internal shocks are generally expected 
to be ``mildly'' relativistic in the fireball 
rest frame, i.e. characterized by Lorentz factor 
$\gamma_i-1\sim1$, since adjacent shells within the wind are expected to
expand with Lorentz factors which do not differ by more than an
order of magnitude. 
We therefore expect results related to particle
acceleration in sub-relativistic shocks \cite{Bland87} 
to be valid for the present
scenario. In particular, the predicted energy distribution of accelerated
protons is $dN_p/dE_p\propto E_p^{-2}$.

Two constraints must be satisfied by
fireball wind parameters in order to allow proton acceleration to
$E_p>10^{20}$~eV in internal shocks \cite{W95a}:
\begin{equation}
\xi_B/\xi_e>0.02\Gamma_{300}^2 E_{p,20}^2L_{\gamma,52}^{-1},
\label{eq:xi_B}
\end{equation}
in order for the proton acceleration time $t_a$ 
to be smaller than the wind expansion time, and
\begin{equation}
\Gamma>130 E_{20}^{3/4}\Delta t^{-1/4}_{10\rm ms},
\label{eq:G_min}
\end{equation}
in order for the 
synchrotron energy loss time of the proton to be larger than $t_a$.
Here, $\Gamma=300\Gamma_{300}$, $\Delta t=10\Delta t_{10\rm ms}$~ms,
$E_p=10^{20}E_{p,20}$~eV, $L_{\gamma}=10^{52}L_{\gamma,52}{\rm erg/s}$
is the $\gamma$-ray luminosity, $\xi_B$ is the 
fraction of the wind energy density which is carried by magnetic field,
$4\pi r^2 c\Gamma^2 (B^2/8\pi)=\xi_B L$, 
and $\xi_e$ is the fraction of wind energy carried by shock
accelerated electrons. 

Eqs. (\ref{eq:xi_B}) and (\ref{eq:G_min}) imply that protons 
may be accelerated in a GRB wind 
to energy $>10^{20}$~eV, provided that  $\Gamma>100$ and that the
magnetic field is close to equipartition with electrons. The former 
condition, $\Gamma>100$, is remarkably similar to that inferred based on
$\gamma$-ray spectra. 
There is no theory at present that allows a basic
principles calculation of the strength of the magnetic field. However, 
magnetic field close to equipartition, $\xi_B\sim1$, is required
in order to account for the observed $\gamma$-ray emission. 

We have assumed in the discussion so far that the fireball is spherically 
symmetric. However, since a jet-like fireball behaves as if it were
a conical section of a spherical fireball as long as the jet opening
angle is larger than $\Gamma^{-1}$, our
results apply also for a jet-like fireball 
(we are interested only in processes that occur when
the wind is ultra-relativistic, $\Gamma\sim300$, prior to 
significant fireball deceleration). For a jet-like wind, $L$ in our
equations should be understood as the luminosity the fireball
would have carried had it been spherically symmetric.

\section{Implications of afterglow observations}

In addition to providing support to the validity of the qualitative fireball 
scenario described in \S2.1, afterglow observations provide 
quantitative constraints on fireball model parameters.
The determination of GRB redshifts implies that
the characteristic GRB $\gamma$-ray luminosity and emitted energy
are $L_\gamma\sim10^{52}{\rm erg/s}$ and 
$E_\gamma\sim10^{53}{\rm erg}$ respectively, 
an order of magnitude higher than the values
assumed prior to afterglow detection.
Afterglow observations also indicate that $\xi_e\sim\xi_B\sim0.1$.
This suggests that the constraint 
(\ref{eq:xi_B}) is indeed satisfied, allowing proton acceleration to
$>10^{20}$~eV.

The observed GRB redshift distribution implies a GRB rate of 
$R_{\rm GRB}\sim10/{\rm Gpc}^3{\rm yr}$ at $z\sim1$. 
The present, $z=0$, rate is less well constrained, since most observed 
GRBs originate at redshifts $1\le z\le2.5$ \cite{GRB_z}. 
Present data
are consistent \cite{GRB_z} 
with both no evolution of GRB rate with redshift, and 
with strong evolution (following, e.g.,
the luminosity density evolution of QSOs or the evolution of
star formation rate), in which 
$R_{\rm GRB}(z=1)/R_{\rm GRB}(z=0)\sim8$.
The energy observed in $\gamma$-rays reflect the fireball
energy in accelerated electrons. If shock accelerated protons and electrons 
carry similar energy, as indicated by afterglow observations,
then the $z=0$ rate of cosmic-ray production by GRBs is
similar to the generation rate of $\gamma$-ray energy, 
\begin{equation}
E^2 (d\dot n_{CR}/dE)_{z=0}\approx 10^{44}\zeta {\rm erg/Mpc}^3{\rm yr},
\label{eq:cr_rate}
\end{equation}
where $\zeta$ is in the range of $\sim1$ to $\sim8$.

\section{Comparison with UHECR observations}

In Fig. 1 we compare the UHECR spectrum,
reported by the Fly's Eye \cite{Bird934}, the Yakutsk \cite{Yakutsk}, 
and the AGASA \cite{Takeda98} experiments, 
with that predicted by the GRB model.
\begin{figure}
\centerline{\psfig{figure=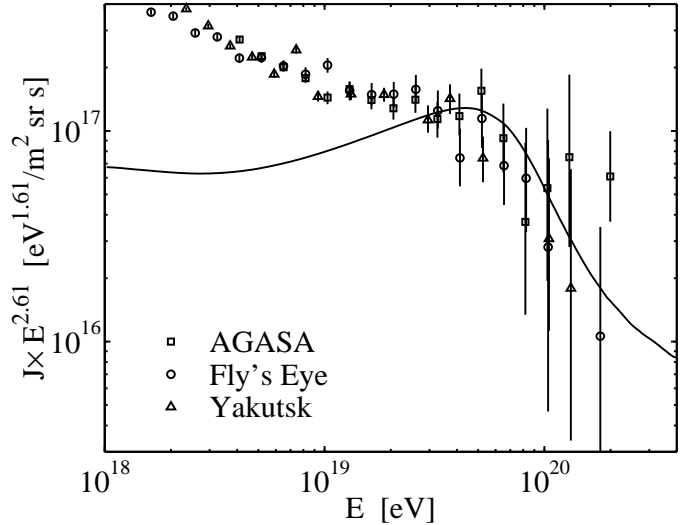,width=3.5in}}
\caption{
The UHECR flux expected in a cosmological model, where high-energy protons 
are produced at a rate $(E^2 d\dot n_{CR}/dE)_{z=0}=0.8\times10^{44}
{\rm erg/Mpc}^3{\rm yr}$ as predicted in the GRB model 
[Eq. (\ref{eq:cr_rate})]
(The flux above $10^{19}$~eV is not sensitive to 
the $z$ dependence of $E^2 d\dot n_{CR}/dE$).
$1\sigma$ flux error bars are shown. The highest energy points are derived
assuming the detected events (1 for Fly's Eye and Yakutsk, 
4 for AGASA) represent a
uniform flux over the energy range $10^{20}$~eV--$3\times10^{20}$~eV.}
\label{fig1}
\end{figure}
The flattening of the cosmic-ray spectrum at $\sim10^{19}$~eV,
combined with the lack of anisotropy 
and the evidence for a change in composition from heavy nuclei at low
energy to light nuclei (protons) at high energy \cite{composition},
suggest that an extra-Galactic source of protons dominates the flux at
$E>10^{19}$~eV. The UHECR flux predicted by the GRB model
is in remarkable agreement with the observed extra-Galactic flux.

The suppression of model flux above $10^{19.7}$~eV is  
due to energy loss of high energy protons
in interaction with the microwave background, i.e. to the 
``GZK cutoff'' \cite{GZK}. 
Both Fly's Eye and Yakutsk data show a deficit in the number of events, 
consistent with the predicted 
suppression. The deficit is, however, only
at a $2\sigma$ confidence level \cite{Nobel_rev}.
The AGASA data is consistent 
with Fly's Eye and Yakutsk results below $10^{20}$~eV.
A discrepancy may be emerging at higher energy, $>10^{20}$~eV, 
where the Fly's Eye and Yakutsk experiments detect 1 event each,
and the AGASA experiment detects 6 events for similar exposure. 

The flux above $10^{20}{\rm eV}$ is
dominated by sources at distances $<40\ {\rm Mpc}$. 
Since the distribution of known astrophysical systems
(e.g. galaxies, clusters of galaxies) is inhomogeneous on scales of
tens of Mpc, significant deviations from model predictions presented
in Fig. 1 for a uniform source distribution are expected above 
$10^{20}{\rm eV}$.
Clustering of cosmic-ray sources leads \cite{CR_clustering}
to a standard deviation, $\sigma$, in the expected number, $N$, of 
events above $10^{20}$ eV, given by 
$\sigma /N = 0.9(d_0/10 {\rm Mpc})^{0.9}$, where $d_0$ is the unknown scale
length of the source correlation function and $d_0\sim10$ Mpc 
for field galaxies.

An order of magnitude increase in the exposure of UHECR experiments,
compared to that available at present, is required to test for
the existence of the GZK
cutoff. Such exposure would allow this test through an  
accurate determination of the spectrum in the energy range
of $10^{19.7}$~eV to $10^{20}$~eV, where the effects of source inhomogeneities
are expected to be small \cite{CR_clustering}. 
Moreover, an order of magnitude
increase in exposure will also allow to determine the source correlation 
length $d_0$, through the detection of anisotropies in the arrival directions
of $\sim10^{19.5}$~eV cosmic-rays over angular scales of 
$\Theta\sim d_0/30$~Mpc \cite{CR_clustering}.

\section{GRB model predictions for planned UHECR experiments}

The rate at which GRBs occur within a distance of $\sim100$~Mpc from Earth,
the distance to which $>10^{20}$~eV proton propagation is limited due to 
interaction with the microwave background, is $\sim1$ per 100~yr. This
rate can be reconciled with the detection of several $>10^{20}$~eV events
over a period of a few years only if
there is a large dispersion, $\geq100{\rm yr}$, in the arrival time of protons 
produced in a single burst. The required dispersion
is likely to result from deflection by random magnetic fields \cite{W95a}. 
A proton of energy $E$ propagating over a distance $D$
through a magnetic field of strength $B$ and correlation length
$\lambda$ is deflected by an angle
$\theta_s\sim(D/\lambda)^{1/2}\lambda/R_L$, which results in a time
delay, compared to propagation along a straight line,
$\tau(E,D)\approx\theta_s^2D/4c\propto B^2\lambda$. 
The random energy loss suffered by $>10^{20}{\rm eV}$ protons 
as they propagate, owing to the 
production of pions, implies that protons observed at Earth 
with given energy have different energy histories along their 
propagation path. Thus, magnetic field deflection results not only in 
a delay, but also in a spread in arrival time of protons of fixed energy, 
comparable to the delay $\tau$.

The current upper bound on the inter-galactic magnetic 
field \cite{IGM}, $B\lambda^{1/2}\le10^{-9}{\rm G\ Mpc}^{1/2}$,
allows a spread $\tau(E=10^{20}{\rm eV},D=100{\rm Mpc})\sim10^5$~yr,
well above the minimum, $\tau\sim100$~yr, required in the 
model. The magnetic field upper bound 
implies an upper bound on the number of GRBs contributing to the
$>10^{20}$~eV flux at any given time, $\sim10^5/100=10^3$. The upper bound
on the number of sources contributing to the flux above $E$ decreases rapidly 
as $E$ increases beyond $10^{20}$~eV, as the propagation distance 
decreases with energy due to the increase in pion production energy loss
rate. This rapid decrease implies that at $E\sim3\time10^{20}$~eV 
there can be only a few sources contributing to the flux at any given 
time \cite{MnW96}.

The GRB model therefore makes a unique prediction \cite{MnW96}: 
The UHECR flux at
energy $E\ge3\times10^{20}$~eV should be dominated by a few sources on the 
sky. These sources should have narrowly peaked energy spectra,
and the brightest sources should be different at different energies.
This is due to the fact that
at any fixed time a given burst is observed in UHECRs only over a narrow
range of energy: If
a burst is currently observed at some energy $E$ then UHECRs of much lower
(higher) energy from this burst will arrive (have arrived) mainly in the 
future (past). 
Testing the GRB model predictions 
requires an exposure 10 times larger than that of present
experiments. Such increase is expected to be provided by the planned
Auger \cite{Auger} detectors.

\section{ $10^{14}$~eV Neutrinos}

Protons accelerated in the fireball to high energy lose energy through
photo-pion production in interaction with fireball photons. 
The decay of charged
pions results in the production of high energy neutrinos \cite{WnB97}. 
The observed energy of a proton, for which the observed 1~MeV photons
are at the threshold of the $\Delta$-resonance, is 
$0.2{\rm\, GeV}^2\Gamma^2/1$~MeV. Typically, the neutrino receives 
$\sim5\%$ of the proton energy. Thus, the typical energy of neutrinos 
resulting from interaction of accelerated protons with GRB photons is\
\begin{equation}
E_\nu^b\approx5\times10^{14}\Gamma_{300}^2{\rm eV}.
\label{Enu}
\end{equation}

The flux normalization is determined by the
efficiency of pion production.
The fraction of energy lost to pion production
by protons producing the neutrino flux above $E^b_\nu$ is 
essentially independent of energy and is given by \cite{WnB97}
\begin{equation}
f_\pi\approx0.2{L_{\gamma,52}\over\Gamma_{300}^4 \Delta t_{10\rm  ms}}.
\label{fpi}
\end{equation}
If GRBs are the sources of UHECRs, 
then using Eq. (\ref{fpi}) and and Eq. (\ref{eq:cr_rate}) with $\zeta\simeq1$,
the expected GRB neutrino flux is \cite{WnB97}
\begin{equation}
E_\nu^2\Phi_{\nu_x}\approx
1.5\times10^{-9}\left({f_\pi\over0.2}\right)\times
\min\{1,E_\nu/E^b_\nu\}
{{\rm GeV}\over{\rm cm}^{2}{\rm s\,sr}},
\label{JGRB}
\end{equation}
where $\nu_x$ stands for $\nu_\mu$, $\bar\nu_\mu$ and $\nu_e$.

The flux of $\sim10^{14}$~eV neutrinos given in Eq. (\ref{JGRB}) implies
that large area, $\sim1{\rm km}^2$, high-energy neutrino telescopes,
which are being constructed to detect 
cosmologically distant neutrino sources \cite{Halzen_review99}, would
observe several tens of events per year correlated in time and
in arrival direction with GRBs. 
Detection of neutrinos from GRBs could be used to
test the simultaneity of
neutrino and photon arrival to an accuracy of $\sim1{\rm\ s}$
($\sim1{\rm\ ms}$ for short bursts), checking the assumption of 
special relativity
that photons and neutrinos have the same limiting speed
to one part in $10^{16}$, and
the weak
equivalence principle, according to which photons and neutrinos should
suffer the same time delay as they pass through a gravitational potential,
to one part in $10^{6}$ (considering the Galactic potential alone).

The model discussed above predicts the production of high energy
muon and electron neutrinos. 
However, if the atmospheric neutrino anomaly has the explanation it is
usually given, oscillation to $\nu_\tau$'s with mass 
$\sim0.1{\rm\ eV}$ \cite{atmo}, then
one should detect equal numbers of $\nu_\mu$'s and $\nu_\tau$'s. 
Up-going $\tau$'s, rather than $\mu$'s, would be a
distinctive signature of such oscillations. 
Since $\nu_\tau$'s are not expected to be produced in the fireball, looking
for $\tau$'s would be an ``appearance experiment.''
To allow flavor change, the difference in squared neutrino masses, 
$\Delta m^2$, should exceed a minimum value
proportional to the ratio of source
distance and neutrino energy. A burst at $100{\rm\ Mpc}$ 
producing $10^{14}{\rm eV}$ neutrinos can test for $\Delta m^2\ge10^{-16}
{\rm eV}^2$, 5 orders of magnitude more sensitive than solar neutrinos.

\end{document}